\documentclass[12pt,a4paper]{article}
\usepackage[utf8]{inputenc}
\usepackage[margin=1.6cm]{geometry}
\usepackage{amsmath}
\usepackage{amsfonts}
\usepackage{authblk}
\usepackage{amssymb}
\usepackage{graphicx}
\usepackage{mathtools}
\usepackage{xcolor}
\usepackage{hyperref}
\usepackage[title]{appendix}
\usepackage[capitalise]{cleveref}
\author[1,*]{Ander Movilla Miangolarra}
\author[1]{Martin Howard}
\title{Theory of epigenetic switching due to stochastic histone mark loss during DNA replication}
\affil[1]{\small Department of Computational and Systems Biology, John Innes Centre, Norwich NR4 7UH, United Kingdom}
\affil[*]{email: ander.movilla-miangolarra@jic.ac.uk}
\begin{document}
\maketitle

\begin{abstract}
How much information does  a cell inherit from its ancestors beyond its genetic sequence? What are the epigenetic  mechanisms that allow this? Despite the rise in available epigenetic data, how such information is inherited through the cell cycle is still not fully understood. Often, epigenetic marks can display  bistable behaviour and their bistable state is transmitted to daughter cells through the cell cycle, providing the cell with a form of memory. However, loss-of-memory events also take place, where a daughter cell switches epigenetic state (with respect to the mother cell). Here, we develop a framework to compute these epigenetic switching rates, for the case when they are driven by DNA replication, i.e., the frequency of loss-of-memory events  due to replication.  We consider the dynamics of histone modifications during the cell cycle deterministically, except at DNA replication, where nucleosomes are randomly distributed between the two daughter DNA strands, which is therefore implemented stochastically. This hybrid stochastic-deterministic approach enables an analytic derivation of the replication-driven switching rate. While retaining great simplicity, this framework can explain experimental switching rate data, establishing its biological importance as a framework to  quantitatively study epigenetic inheritance.

\end{abstract}

\section{Introduction}
During the $20^\textrm{th}$ century, crucial breakthroughs were achieved to understand the role of nucleic acids in biological inheritance. At the beginning of the $21^\textrm{st}$ century, the technology to sequence and assemble the DNA of the most studied organisms was deployed. Nevertheless, our understanding of how this genetic material is regulated is still in its infancy, despite the vast amount of genetic and epigenetic data experiments can generate \cite{encode2020expanded}. Reasons for our poor understanding of genetic regulation, especially in eukaryotes, include the combinatorial complexity of the interactions among genes and their products and the lack of established mechanistic and quantitative frameworks to understand vast amounts of  data.

One of the most intriguing problems of gene regulation is how cells can inherit gene regulatory patterns from their ancestors. Especially in higher eukaryotes, but also in unicellular cases, different RNA expression patterns can be reliably inherited through the cellular lineage, despite all cells sharing the same underlying DNA. Often, methylation of DNA  at CpG sites carries the transcriptional information required but, in many other cases, this information is carried  in the form of post-translational modifications (PTMs) of nucleosomes---histone octamers around which the DNA wraps in the eukaryotic nucleus \cite{Schlissel2019,groth2023symmetric}.

In the case of nucleosomes, since there are insufficient parental copies to fully occupy both daughter DNA strands after replication, they are newly synthesised by the cell in advance of S phase \cite{armstrong2021replication}. This implies that, after DNA replication, any histone PTMs carried within the parental cell's DNA  will be diluted in the daughter cells among the newly synthesised nucleosomes (devoid, in general, of epigenetic marks). This (on-average) two-fold serial dilution of histone marks begs the question of how  such a form of memory can be stably inherited over many generations \cite{ANNUNZIATO2005}. However, we know that the so-called `reader-writer' enzymes are at the core of the solution to this enigma. This  class of enzymatic complexes is capable of binding to particular histone PTMs (`reading') and spreading these same PTMs to nearby nucleosomes through their catalytic activity (`writing') \cite{Allshire2018Ten}. In this way, the histone PTMs lost at replication can be recovered and a long-lasting epigenetic identity can be established, which can be inherited through the cell cycle \cite{Briffa2024,Irish2024,Jolly2023}.

Nevertheless, loss-of-inheritance events do take place and the cause of these switching events remains uncertain. In this work, we consider a classification for the loss of inheritance in terms of the underlying cause:  either \textit{replication-driven}, where the perturbation  due to DNA replication and the consequent dilution of histone PTMs causes the switching; or driven by the inherent stochasticity of other cellular processes (e.g., noise in read/write biochemical reactions). Leveraging this classification, we build a mathematical framework for replication-driven epigenetic switching. This framework is  based on deterministic dynamics of histone PTMs during the cell cycle, coupled to a stochastic perturbation at S phase due to DNA replication.

Previous modelling efforts have shown that the feedback mechanisms of these `reader-writer' systems can yield the bistability and epigenetic memory  observed in experiments over many cell cycles \cite{marenduzzo2020competition,berry2017slow,Briffa2024,dodd2007theoretical, Jolly2023,Lovkvist2021Protein,michieletto2016polymer, Nickels2021,owen2023design}.  However, these works were heavily based on numerical simulations of stochastic systems and, beyond the read-write feedback and the appearance of bistable behaviour, it has been unclear, in general, what fundamental principles govern the faithful inheritance of transcriptional information and which sources of noise could destabilise it. Here, we sought a more analytically tractable approach, which combines deterministic differential equation modelling throughout the cell cycle (which could be visualised as an epigenetic landscape) and stochastic perturbations due to DNA replication. This approach enabled us to derive simple relations for the switching rates due to replication and compare them with experimental data.

 Noteworthy in this line of research is  work by Micheelsen and colleagues \cite{micheelsen2010theory},  also aimed at obtaining analytical results for this type of system (see also \cite{jost2014bifurcation}, for a similarly interesting approach). However, their approach was hindered by the complexity and stochasticity involved, which we circumvented by simplifying the problem to address replication-driven switching only. Moreover, landscape approaches have already been proposed to account for transcriptomic and epigenetic data (e.g. \cite{siggia2012geometry,saez2022dynamical,teschendorff2021statistical}), but this has usually been done in a phenomenological way, to account for heterogeneity and cell fate decisions. Here, instead, we take a bottom-up approach, placing more emphasis on the mechanisms and resolving the dynamics at the scale of the cell cycle, by taking explicitly into account the effects of replication on the epigenetic information.

Here, we illustrate this procedure to compute replication-induced switching for two bistable epigenetic models, which have been used to model heterochromatin in two different yeast species. The first one is a simple two-dimensional   model that describes epigenetic switching in the mating region in fission yeast \cite{dodd2007theoretical}, whose simplicity allows for an analytical approximation of the replication-driven switching rate. The second model is a more complex one for the \textit{HMR} locus in budding yeast, which we reduce to a single dimension and successfully compare to experimental data. In the last section, we discuss how this method may be generalised to any model, in order to compute its replication-driven switching rates.

\section{Mathematical model for epigenetic switching in fission yeast} \label{fission}

The first model of epigenetic inheritance  we consider is based on that of Ref. \cite{dodd2007theoretical}. It is a simple and convenient model to illustrate replication-driven switching and has also been extensively used to theoretically interrogate  epigenetic inheritance systems for almost two decades \cite{Briffa2024,Nickels2021}. The model describes histones as either acetylated $A$, unmodified $U$, or methylated $M$  and takes into account the transitions between them: $A \leftrightharpoons U \leftrightharpoons M$ (where $A$ is associated with an active locus and $M$ with a silent one). It is a particularly well-suited model for heterochromatin in fission yeast, where these states can be related to the PTM state of histone H3 at lysine 9 (H3K9). 

In the spirit of replication-driven transitions, we neglect all other sources of noise in the dynamics during the cell cycle.  Our approach hinges on this assumption, which allows for an extensive analysis of the system. We emphasise that there is no guarantee that in every system this will be a good approximation of the dynamics. Thus, the analysis that follows will only be valid in cases where the dilution of histone PTMs due to DNA replication is the largest source of noise.

 Within this approximation, the deterministic dynamics of the concentrations of acetylated (or methylated) histones during the cell cycle in a given locus of interest, $c_a$ (or $c_m$), can be described as
\begin{align}
\frac{d c_a}{dt}&=(1-c_a-c_m)(1+k_c c_a)-c_a k_d (1+k_c c_m) \label{eqa0} \\
\frac{d c_m}{dt}&=(1-c_a-c_m)(1+k_c c_m)-c_m k_d (1+k_c c_a) \label{eqm0},
\end{align}
where $k_c, k_d>0$, the time $t$ has been rescaled to absorb the methylation/acetylation basal rate constant, and the concentration of unmodified histones is $c_u=1-c_a-c_m$ due to the normalisation. For simplicity, the system has been chosen to be symmetric with respect to a $c_a \leftrightharpoons c_m$ rotation, although, in practice, it is unlikely that acetylation and methylation are exactly identical processes. 

\begin{figure}
\includegraphics[width=\textwidth]{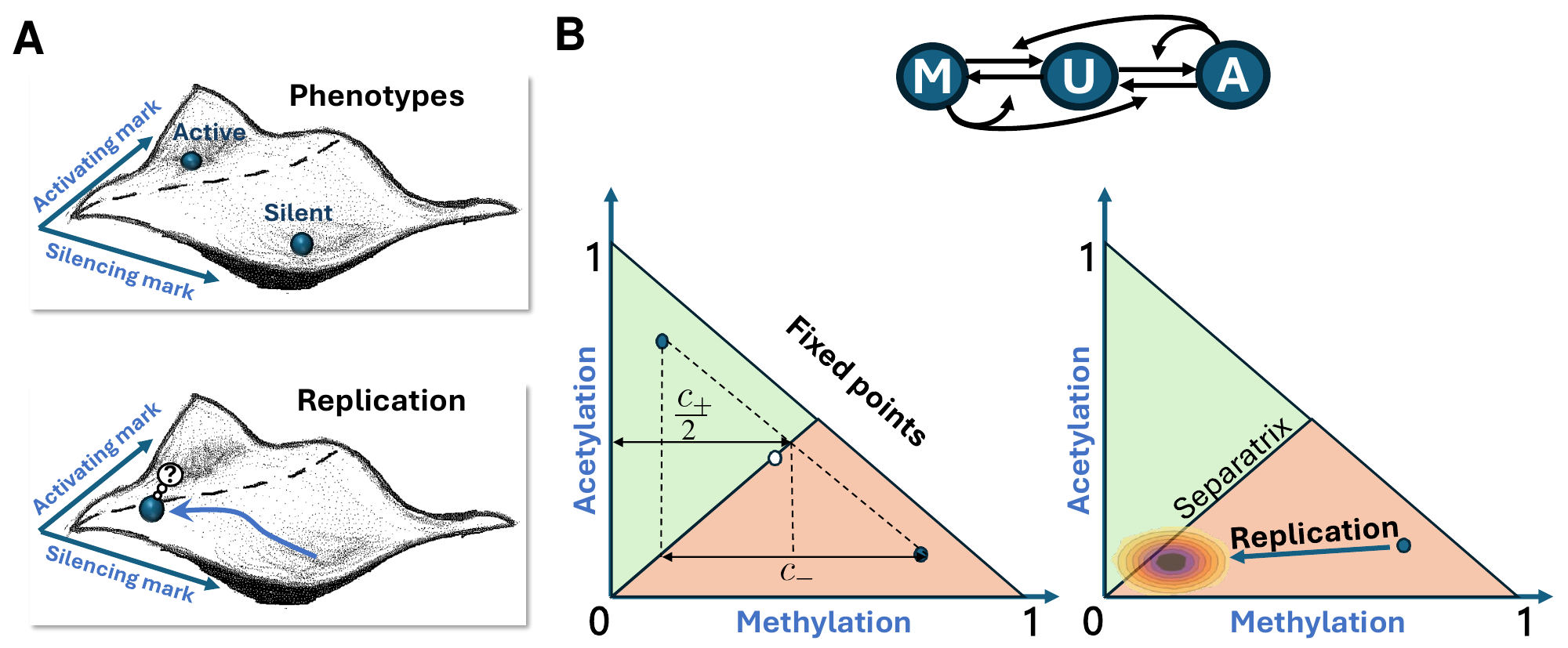}
\caption{Landscape analogy of epigenetic regulation and inheritance. A) Illustration of the epigenetic landscape. The minima of the epigenetic landscape, whose coordinates are the abundances of specific epigenetic marks, correspond to the observable silent/active phenotypes (top). DNA replication perturbs these states by introducing unmarked nucleosomes and pushing the cell up the landscape (bottom). B) Mathematical equivalent of the landscape. The phenotypes are the fixed points of a dynamical system (left). Replication kicks the system away from the fixed point by a random distance (right), from where it will evolve in the following cell cycle. Switching occurs if the stochastic perturbation due to replication brings the system into the  basin of attraction of the other fixed point, i.e., crosses the separatrix. Top: diagram of the regulatory network considered. \label{Fig1}}
\end{figure}

In every transition term of the model, there is a basal rate and a catalytic rate (which is parametrised by $k_c$). The latter  depends on the recruitment of relevant enzymes by nucleosomes present at the locus and can be related to the activity of `reader-writer' enzymes, since the rate involves a product of the concentrations: its substrate (the histone they `write' to) and the histone PTM to which they bind (the histone PTM they `read'). Note that the catalytic terms represented by $k_dk_c$ terms do not correspond to the usual read-write feedback as they erase the opposite mark to the one they bind (evidence for this kind of feedback can be found in Refs. \cite{wang2013epe1,yang2016flc}). For appropriate parameters, these catalytic rates embody sufficiently strong feedbacks to stably maintain two fixed points, representing silenced or expressed genes. Conversely, the basal rates represent a degree of noise or cross-talk in the system. This could be either due to the action of freely diffusing histone modifiers, or enzymes bound at other locations that occasionally contact the locus, or enzymes recruited through other means (e.g., histone acetyltransferases activated by trancription factors \cite{ortega2018transcription}). The model is schematically shown in Fig. \ref{Fig1}B, top.


In the rest of this section we first analyse the deterministic dynamics of the model and afterwards we add the stochastic component due to replication.

\subsection{Fixed points}

The system formed by \cref{eqa0,eqm0} has four fixed points, denoted by the superscript $\ast$, but only three of them in the positive $c_a$ and $c_m$ quadrant (Fig. \ref{Fig1}B):
\begin{align}
c^{*}_{a,1}=&c^{*}_{m,1}=\frac{1}{k_d+2} \label{ss01} \\
c^{*}_{a,2}=&(c_+ + c_-)\frac{1}{2}, \quad c^{*}_{m,2}=(c_+ - c_-)\frac{1}{2} \label{ss02} \\
c^{*}_{a,3}=&(c_+ - c_-)\frac{1}{2}, \quad c^{*}_{m,3}=(c_+ + c_-)\frac{1}{2} \label{ss03},
\end{align}
where
\begin{align}
c_+&=1-\frac{k_d}{k_c} \label{c+}, \\
c_-&=\frac{1}{k_c} \sqrt{(k_c-k_d)^2-4}. \label{c-}
\end{align}
Local stability can be assessed by linear stability analysis, see Appendix \ref{LSA}. The first fixed point is locally stable if $k_c<k_d+2$, and a saddle point otherwise. With respect to the phase space $(c_a, c_m)$, it is always stable in the $(1,1)$ direction, but can be unstable in the $(-1,1)$ direction. Thus, for $k_c>2+k_d$, the fixed point defined by \cref{ss01} is unstable in favour of the second and third fixed points, which are now stable (see Fig. 2A for examples of trajectories in phase space).

Using a landscape as a metaphor for the dynamical system of \cref{eqa0,eqm0}, the fixed points would be the local minima of the landscape. When evolving in time, the system would descend through the landscape, from the initial condition to the fixed point (see Fig. 1).

\subsection{Separatrix}

We have seen there are two stable fixed points in the dynamical system, but we would like to know, within the allowed  phase space, which initial conditions would take us to one fixed point and which ones would take us to the other fixed point, i.e. the basins of attraction of each fixed point. In 2 dimensions, the separatrix is a line that separates the basins of attraction. It originates at the saddle point (if it exists) and is propagated across the phase space by the evolution of the dynamical system along the stable eigenvector of the saddle point \cite{murray2007mathbio}. 

In this case, for $k_c>k_d+2$, the separatrix passes through the point  $c^{*}_{a,1}=c^{*}_{m,1}=\frac{1}{k_d+2}$, along the $v_+=(1,1)$ direction. The dynamical system along this line is symmetric, implying
\begin{equation}
\label{sep_cond}
\frac{dc_a}{dt}=\frac{dc_m}{dt} \rightarrow \frac{d c_a}{dc_m}=1.
\end{equation}
Then, the separatrix is the line $c_a=c_m$, which divides the two basins of attraction of the dynamical system \cref{eqa0,eqm0} (Figs. \ref{Fig1}B and \ref{Fig2}A). In the landscape picture, the separatrix is the ridge between the valleys, separating their basins of attraction.

\subsection{Replicative dilution} \label{Rep_dilution}

	Given that  the dynamics during the cell cycle were modelled deterministically, the only possibility for a locus in one state to switch to another one is due to a large fluctuation at replication.
	
	The dynamical system considers the fraction of histones with a particular PTM as a continuum, which is accurate for large nucleosome numbers. In this same limit, the binomial distribution,  used to model the random inheritance of nucleosomes to daughter DNA strands, can be  approximated by a Normal distribution:
	\begin{equation}
	\mathcal{B}(n,p) \sim \mathcal{N} \left(n p, n p(1-p) \right),
	\end{equation}
	where $n$ is the number of nucleosomes with a given PTM and $p$  is the probability of a given strand  to inherit a given nucleosome. Typically, if there is no bias in the DNA replication machinery, a standard assumption is $p=0.5$, implying that a given nucleosome can be inherited with equal probability by either DNA strand \cite{ANNUNZIATO2005,Schlissel2019}. Equivalently, there is a 50\% probability that a given nucleosome will not be inherited onto a given daughter strand and, thus, in the daughter strand, the nucleosomal location will be occupied by a newly synthesised nucleosome (unmodified, in this case).
	
	In the 3-state epigenetic model we propose, $A$ or $M$ marked nucleosomes would be either inherited or replaced by unmodified nucleosomes. According to the Normal approximation, if before replication the locus had arrived at one of the steady states $(N c_a^*, N c_m^*)$ (where $N$ is the total number of nucleosomes), the probability distribution for the locus to inherit $(N c_a^i, N c_m^i)$ nucleosomes after replication is
	\begin{equation}
	P_R(c_a^i, c_m^i \vert c_a^*, c_m^*)=\frac{2N}{\pi \sqrt{c_a^* c_m^*}} \exp \left[-2N\frac{(c_a^i-c_a^*/2)^2}{c_a^*}-2N\frac{(c_m^i-c_m^*/2)^2}{c_m^*} \right].
\end{equation}	
The $(c_a^i, c_m^i)$ pair constitutes the initial condition of the dynamical system for the next generation, and, hence, it determines  which fixed point the locus will evolve towards. In this sense, replication can be seen as a strong perturbation that pushes the system up the landscape and if it pushes it far enough (beyond the separatrix) it will cause the switching, see Fig. 1. 

We note that the assumption that the system reaches the fixed point before DNA replication is violated in certain cases \cite{Goodnight2020sPhase}, but we expect it to be a good approximation for many histone PTMs, especially for acetylation, which is thought to turn over on a timescale of tens of minutes \cite{Waterborg2001}. However, the H3K9me3 turnover timescale is usually of the order of the cell cycle and certain other PTMs, such as those catalysed by Polycomb Repressive Complex 2, will take longer than a cell cycle to settle into a fixed point \cite{alabert2015two,zee2010vivo}. We refer the interested reader to Ref. \cite{zerihun2015effect}, where the consequences of this assumption being violated were numerically studied.

A further assumption implicit in this analysis is the fact that both histone H3 copies within a given nucleosome are marked with the same PTM. While this is predominantly the case in very polarised scenarios, where most H3 histones are either acetylated or methylated, in cases where the fixed points are closer to the separatrix there could be nucleosomes with mixed PTMs, whose existence, in this model, we are neglecting. 


\subsection{Switching rate} \label{switching}

In a system that has sufficient time to reach the fixed point during each cell cycle and whose separatrix is simply $c_a=c_m$, the switching rates are given  by 
\begin{equation}
\label{num_int}
S_1=\int\limits_{\substack{0\le c_a^i \le c_m^i \le 1 \\ \textrm{s.t. } c_a^i + c_m^i \le 1}} P_R (c_a^i, c_m^i  \vert c_{a,2}^\ast,  c_{m,2}^\ast ) dc_a^i dc_m^i,
\end{equation}
where $S_1$ is the rate of the transition from a high acetylation to a high methylation state. $S_2$, corresponding to the opposite transition, is equal to $S_1$ ($S_2=S_1=S$) given the symmetry of the model.

 \begin{figure}
 \center
 \includegraphics[width=0.75\textwidth]{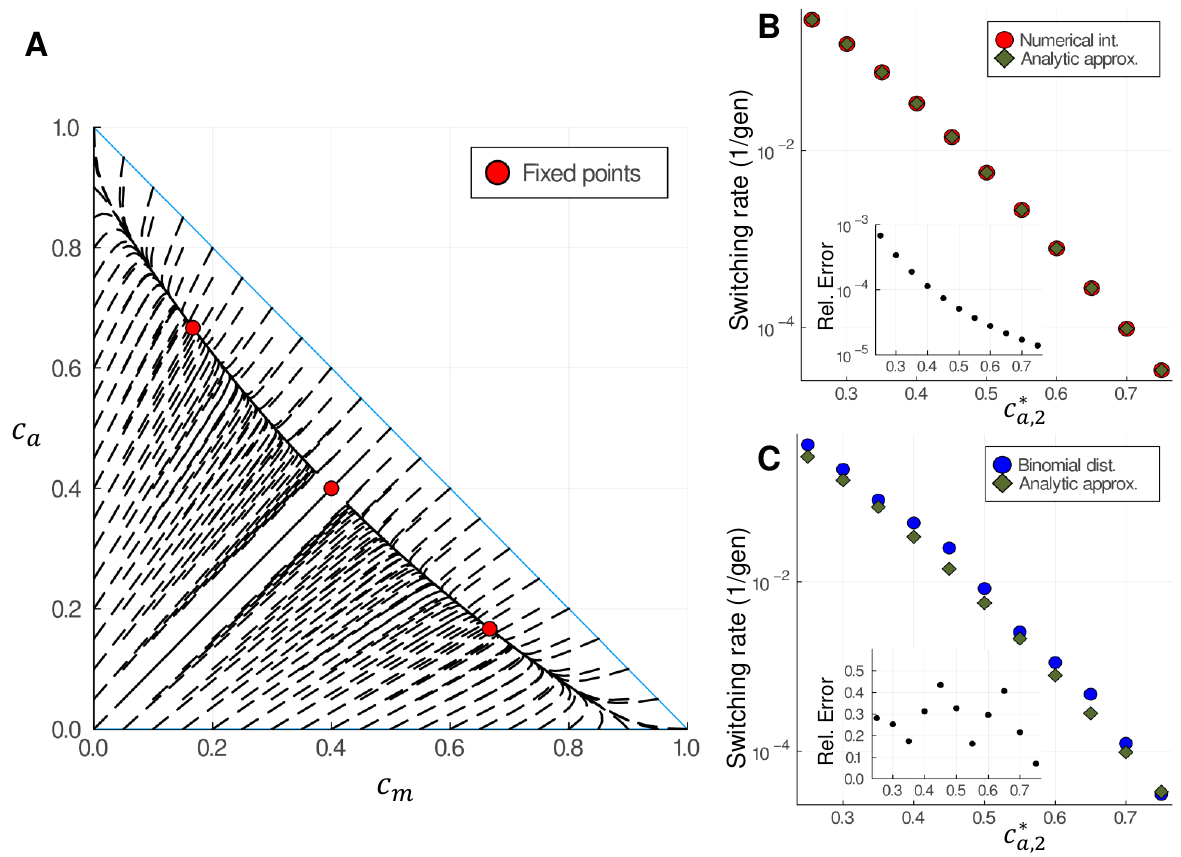}
\caption{A) Phase diagram of the model for $k_d=0.5$ and $k_c=3$, with the dashes representing the flow on the phase portrait and the red circles, the fixed points. The separatrix is the line $c_m=c_a$, which passes through the unstable fixed point. B) Switching rates for the $c_m=c_a$ separatrix, with $c_{m,2}^\ast$ fixed at 0.2 and varying $c_{a,2}^\ast$, for both the analytic approximation \cref{switching_rate} and the numerical integration of \cref{num_int}. Inset: Relative error of the analytical approximation with respect to the numerical integration. C) Same as B, but comparing the switching rates obtained from the analytical approximation \cref{switching_rate} and the binomial distribution with $N=50$. While the relative error is of the order of 20-30\%, it is still small enough not to obscure the overall trend, which spans over more than two orders of magnitude.  \label{Fig2}}
\end{figure}

\subsection{Analytical approximation}

We can compute the integrals if we extend the integration domains:
\begin{equation}
\label{int_approx}
S \simeq \int \limits_{-\infty < c_a^i \le c_m^i < +\infty} P_R (c_a^i, c_m^i  \vert c_{a,2}^\ast,  c_{m,2}^\ast ) dc_a^i dc_m^i.
\end{equation}
Note that the domains have been increased substantially (infinitely!) but most of the distribution should fall within the original integration domain, making the error of the approximation very small (see Fig. \ref{Fig2}B, inset). Then, the integral can be evaluated analytically (see Appendix \ref{Integrals}):
\begin{equation}
\label{switching_rate}
S\simeq \frac{1}{2} \left[ 1-\textrm{erf} \left( \sqrt{\frac{N}{2}}\frac{c_-}{\sqrt{c_+}} \right)\right].
\end{equation}
This equation for the switching rate can be interpreted as follows: $c_-$ is proportional to the distance of either fixed point to the separatrix (the normal distance is $ c_-/\sqrt{2}$), explaining why the switching rate decreases with increasing $c_-$. Indeed, as $c_- \to 0$, $S \to 1/2$, since in this limit, the fixed point is located on the separatrix. In addition, a factor of $\sqrt{N/c_+}$ is present within the error function, which is related to the variance of the Normal distribution: the larger the number of nucleosomes the smaller the variance (in terms of fraction of nucleosomes) and the smaller the switching rate. 

While the approximation of \cref{num_int} by \cref{int_approx} is very good (see Fig. 2B), approximating the binomial distribution by a Gaussian is not as accurate. To estimate the error introduced by this approximation, we rounded the continuous value of the fixed point to the nearest integer (that is $(N c_a^*, N c_m^*)\simeq(N_a^*, N_m^*)$, with $N_a^*$ and $N_m^*$ integers). With the integer values of the fixed points, we computed numerically the probability that replication described by a binomial distribution pushes the system out of its basin of attraction (i.e., summed the probability distribution over the other basin of attraction, including the $N_a=N_m$ case). This comparison can be seen in  Fig. 2C for $N=50$, which systematically overestimates the switching rate due to the inclusion of the $N_a=N_m$ case. 

Finally, it is noteworthy that these results do not depend directly on the dynamical model for the PTMs, \cref{eqa0,eqm0}; they only depend on the fixed points and the separatrix, and the fact that we are studying replication-induced switching. More generally, if the separatrix was a line $c_a=\alpha \, c_m +\beta$ (for $\alpha>0$), the integral for the switching rates can be analogously evaluated to
\begin{equation}
\label{switch_general}
S\simeq\frac{1}{2} \left[ 1- \textrm{erf} \left( \sqrt{\frac{N}{2}}\frac{ c_{a,2}^\ast - 2 \beta - \alpha c_{m,2}^\ast}{\sqrt{c_{a,2}^\ast + \alpha^2 c_{m,2}^\ast}}\right) \right],
\end{equation}
see Appendix \ref{Integrals} for details. In cases where the separatrix can be approximated (at least in the relevant region of phase space) by a straight line, \cref{switch_general} will be a valid approximation of the replication-driven switching rate. Thus, the analysis presented here is a general feature of two-dimensional epigenetic systems  and not a characteristic of a particular dynamical system, and is therefore easily generalised to other numbers of dimensions, as done in Section \ref{quantcomp} for the uni-dimensional case.

\subsection{Asymptotic expansion}
With everything else constant (in intensive parameters, i.e. independent of the size of the system), \cref{switching_rate} predicts that the switching rates due to replicative dilution should decrease with increasing $N$ as
\begin{equation}
S \simeq \frac{1}{2}[1- \textrm{erf} (a \sqrt{N})] \sim \frac{e^{-a^2 N}}{2 a \sqrt{ \pi N}},
\end{equation}
where the asymptotic approximation is only valid for small switching rates and $a$ is a constant. Following an alternative approach, in Ref. \cite{micheelsen2010theory}, transitions due to noise in the biochemical reactions controlling epigenetic marks (the other source of stochasticity mentioned in the introduction) were studied, and a similar exponential scaling was found: $S \propto [N \textrm{exp}(N f)]^{-1}$, $f$ being a function related to the biochemical network. Asymptotically, which of these two rates is larger depends on the parameter within the exponential ($a^2$ and $f$), the smallest of which corresponds to the largest contribution to the switching rate. 

\section{Budding yeast model and comparison with experimental data} \label{quantcomp}

In the previous section, we analysed a simple model of heterochromatin and obtained analytical approximations for the replication-driven switching rates. To bring the theory closer to experiments, we now analyse a model for the epigenetic switching of heterochromatin in budding yeast. This model is more complex, which makes it tractable only numerically, but enables a comparison between theory and experiments.

\subsection{Model}

The \textit{HMR} locus is a genetic region in chromosome III of budding yeast containing two mating type genes and flanked by silencer elements. These silencer elements recruit the SIR family of proteins, which contains a histone deacetylase, Sir2, and a protein that binds to unmodified (not acetylated) nucleosomes, Sir3. Based on this read-write feedback, with Sir2 and Sir3 interacting to form the Sir complex, the locus is silenced \cite{rusche2003rev}. In a certain genetic background, the locus is bistable and its transcriptional state is inherited through the cell cycle.

Consequently,  the relevant post-translational modification is acetylation at H4K16. The budding yeast model considers the fraction of unmodified ($u$), Sir-bound ($s$) and acetylated histones ($a=1-u-s$) and their dynamics are given, in the reduced one-dimensional model, by:
\begin{equation}
\label{eq_csir}
\frac{\textrm{d} u}{\textrm{d} t}=-k_{sas2} u  +k_{sil} [1-u-s^\ast(u)] + \frac{[1-u-s^\ast(u)] s^\ast(u) k_{sir2}}{\frac{V_{max}}{1+(V_{max}-1)[u+s^\ast(u)]}},
\end{equation}
where $s^\ast(u)$ is the fraction of Sir-bound nucleosomes for a given $u$ and  takes the value
\begin{equation}
\label{eqsss_mt}
s^\ast(u)=\frac{u^2 k_D c_\textrm{SIR} (u-1-V_{max} u)}{u^2 k_D c_\textrm{SIR} (V_{max}-1)-2V_{max}}.
\end{equation}
These equations are derived in Appendix \ref{App_SIR} from the stochastic model of Ref. \cite{Miangolarra2024twoway}. Briefly, $k_{sas2}$ is the rate constant for  basal acetylation  (in budding yeast mostly due to the activity of the histone acetyltransferase Sas2), $k_{sil}$ is the deacetylation rate constant related to the effect of the silencer elements, and $k_{sir2}$ is the reaction rate constant for Sir2 deacetylation. The last term of \cref{eq_csir} includes non-linearities due to the read-write feedback and the fact that the Sir3 proteins form dimers (modeled by the binding by dimerisation characteristic constant $k_D$). In addition, there is a factor incorporated into the dimerisation and deacetylation processes due to changes in the compaction of the locus driven by deacetylation of histones (parameterised by $V_{max}$, as the volume factor between the most expanded and most compacted configurations). Finally, $c_\textrm{SIR}$ is the concentration of Sir proteins and the parameter that will be varied when comparing to data. For more details regarding the model, see Appendix \ref{App_SIR} and Ref. \cite{Miangolarra2024twoway}.

Note that, for the relevant parameter regime and region of phase space, \cref{eqsss_mt} a is monotonically increasing function for $s^\ast$ in terms of $u$ (see Fig. \ref{FigA1}). Intuitively, the more unmodified nucleosomes available for the Sir proteins to bind, the more Sir-bound nucleosomes there will be and the more compact the locus will be, further enhancing the Sir binding dynamics.

With appropriate parameters, the dynamical model defined by \cref{eq_csir,eqsss_mt} has three fixed points, the middle one being unstable, which limits the basins of attraction of the stable fixed points. Thus, it plays the role of the separatrix in two dimensions. The positions of these fixed points, together with the PTM state of nucleosomes inserted into chromatin at replication, are all the information needed to compute the replication-driven switching rates, as shown below.

\subsection{Replication and switching rate}

In budding yeast, newly-synthesized histones are rapidly acetylated by Sas2 \cite{Shia2005CharacterizationSAS} and, thus, at replication, the fraction of deacetylated nucleosomes (including Sir-bound ones) in chromatin is diluted and acetylated nucleosomes are inserted. As before, we assume that each deacetylated nucleosome is lost with probability 0.5, in favour of an acetylated one. We further assume that all Sir proteins dissociate from nucleosomes at DNA replication (due to interactions with the replication machinery). Hence, before replication, at each fixed point $j$ there are $\tilde{u}^\ast_j=u^\ast_j+s^\ast(u^\ast_j)$ deacetylated nucleosomes, which are diluted to $u_\textrm{ar}$ deacetylated nucleosomes after DNA replication, with probability
\begin{equation}
\label{1DNormal}
P(u_\textrm{ar}\vert \tilde{u}^\ast_\textrm{sil.})=\sqrt{\frac{2 N}{\pi \tilde{u}^\ast_\textrm{sil.}}} e^{-\frac{2 N}{\tilde{u}^\ast_\textrm{sil.}}(u_\textrm{ar}-\tilde{u}^\ast_\textrm{sil.}/2)^2},
\end{equation}
given by the Normal approximation. Immediately after replication, there are $u_\textrm{ar}$ unmodified (and unbound) nucleosomes to which the Sir proteins start to bind, until they equilibrate at $s^\ast(u)$ given by \cref{eqsss_mt}, subject to the constraint $u_\textrm{ar}=s^\ast(u)+u$. Given that $s^\ast(u)$ is monotonically increasing with $u$, for each $u_\textrm{ar}$ there will be a single $u$ and $s^\ast(u)$. Thus, from the silent stable fixed point, if after replication $u_\textrm{ar}<\tilde{u}^\ast_\textrm{unst.}$, then $u<u^\ast_{unst.}$; implying that the limit of the basin of attraction has been crossed and the system will switch to the active fixed point. 

Based on these considerations, the replication-driven silencing loss rate can be approximated by 
\begin{equation}
\label{1Dintegral}
S\simeq \int_{-\infty}^{\tilde{u}^\ast_\textrm{unst.}} P(u_\textrm{ar}\vert \tilde{u}^\ast_\textrm{sil.}) du_\textrm{ar}, 
\end{equation}
Note that, as in Section \ref{fission}, we have taken the limits of integration of \cref{1Dintegral} to $-\infty$. Together with \cref{1DNormal}, we have that
\begin{equation}
\label{SIR_switching}
S\simeq  \frac{1}{2}\left( 1+ \textrm{erf}\left[\sqrt{\frac{2 N}{\tilde{u}^\ast_\textrm{sil.}}}\left(\tilde{u}^\ast_\textrm{unst.}-\frac{\tilde{u}^\ast_\textrm{sil.}}{2}\right)\right]\right).
\end{equation}
Note that, in the \textit{HMR} case, $N=12$, which makes the Normal approximation less accurate than in other cases, yet enough for practical purposes. If $\tilde{u}^\ast_\textrm{unst.}-\tilde{u}^\ast_\textrm{sil.}/2<0$ the evaluation of the error function will be negative and the switching rate will be smaller than 0.5. If 
\begin{equation}
\Bigg \vert \sqrt{\frac{2 N}{\tilde{u}^\ast_\textrm{sil.}}} \left(\tilde{u}^\ast_\textrm{unst.}-\frac{\tilde{u}^\ast_\textrm{sil.}}{2}\right) \Bigg\vert\gg 1,
\end{equation}
 then, for $\tilde{u}^\ast_\textrm{unst.} \ll \tilde{u}^\ast_\textrm{sil.}/2$, the switching rate tends to 0; and for $\tilde{u}^\ast_\textrm{unst.} \gg \tilde{u}^\ast_\textrm{sil.}/2$, it tends to 1, as expected.

\begin{figure}
\center
\includegraphics[width=\textwidth]{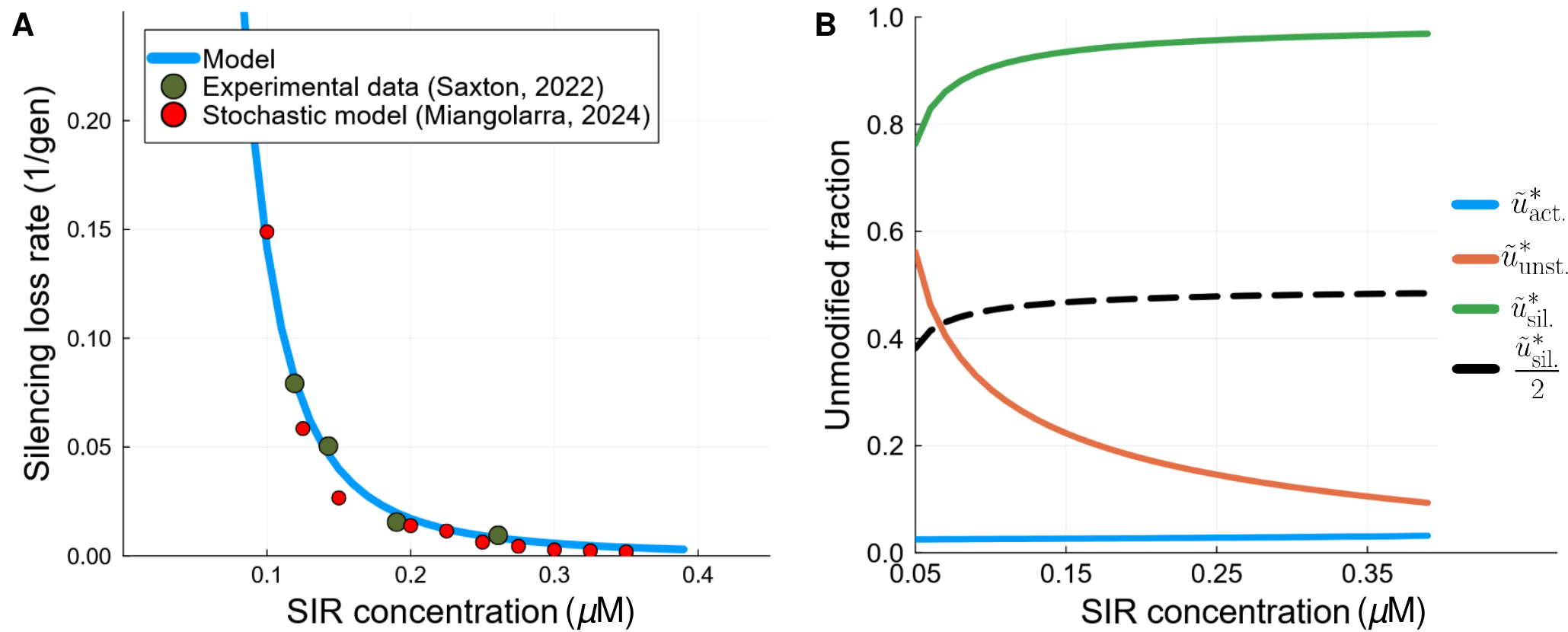}
\caption{A) Comparison with fully stochastic simulations and experimental data from the budding yeast \textit{HMR} locus. The silencing loss rates for various silencing protein (SIR protein) concentrations were measured in Ref. \cite{Saxton2022} and  are compared here to the replication-driven switching rates \cref{SIR_switching} and to the entirely stochastic model from Ref. \cite{Miangolarra2024twoway}. For the replication-driven switching rates, the fixed points of \cref{eq_csir} are obtained numerically for each value of    $c_\textrm{SIR}$ (see also Fig. \ref{Fig4}B), which then enables the computation of the replication-driven switching rate via \cref{SIR_switching}. The agreement is good, both with the experimental data and the fully stochastic simulations, implying that our framework is an accurate and useful description of epigenetic switching. B) For each value of  $c_\textrm{SIR}$, the values of $\tilde{u}^\ast$ are given, for each of the three fixed points (corresponding to the three solid lines). The black dashed line is half of the unmodified fraction of the silent state (i.e., the average level of unmodified nucleosomes after DNA replication in the silent state). The difference between the dashed black line ($\tilde{u}^\ast_{\textrm{sil.}}/2$) and the orange solid line ($\tilde{u}^\ast_{\textrm{unst.}}$) controls the silencing loss rate. \label{Fig4}}
\end{figure}

\subsection{Comparison with experimental data}

We now seek to quantitatively compare model outputs with relevant experimental data to emphasise the biological relevance of the model. Loss-of-inheritance events have been quantitatively studied at this particular locus in budding yeast \cite{Saxton2019,Saxton2022}, enabling comparison between experiments and the model.

Given a complete set of parameters of the model, the replication-driven switching rates are fully specified by the fixed points of \cref{eq_csir}, which are inserted into \cref{SIR_switching} to derive the replication-driven switching rate. Taking $k_{sir2}$ and $k_{sas2}$ as fitting parameters, one can obtain a very good approximation of the experimental switching rates at the \textit{HMR} locus for varying concentrations of the Sir proteins ($c_\textrm{SIR}$), see Fig. \ref{Fig4}A. For details on the fitting procedure, and the specification of the rest of the parameters, see Appendix \ref{budding_numerics}. The quantitative comparison with the experimental data is good, giving support to the replication-induced switching hypothesis. As $c_\textrm{SIR}$ increases, the silencing loss rate decreases as expected, since the Sir proteins are silencing proteins. However, mathematically, this is due to the increasing difference between the unmodified fraction  of the silent steady state and that of the unstable steady state, see Fig. \ref{Fig4}B. In fact, the replication-driven switching rate depends on the difference between $\frac{\tilde{u}^\ast_\textrm{sil.}}{2}$ and $\tilde{u}^\ast_\textrm{unst.}$, as this specifies how much the stochastic process of replication has to deviate from its mean in order to cause the switching. If  $\frac{\tilde{u}^\ast_\textrm{sil.}}{2}-\tilde{u}^\ast_\textrm{unst.}<0$ (as is the case for $c_\textrm{SIR} \lesssim 0.06$, see Fig. \ref{Fig4}B), then $1\le S < 0.5$. Otherwise, $0.5\le S \le 0$, as is the case for higher Sir proteins concentration, tending to 0 for high Sir protein concentrations and large $N$.

In addition, in Fig. \ref{Fig4}A, the results of replication-induced switching are also compared with the results obtained in Ref. \cite{Miangolarra2024twoway} for the fully stochastic model. The agreement between the fully stochastic model and the replication-driven approach implies that, in this model, the overwhelming majority of the epigenetic switching events are induced by DNA replication.  In fact, in Ref. \cite{Miangolarra2024twoway} we had already found in the entirely stochastic version of this model that switching driven by replication is prevalent when switching from a silent  to an active state, but here we have developed a procedure that allows to compute these replication-driven switching rates without requiring stochastic simulations. Altogether, the results from Fig. \ref{Fig4} validate the replication-driven framework from a mathematical and biological standpoint.

\section{Discussion}

In this paper, we have introduced a framework to study epigenetic memory inheritance that only considers stochastic loss-of-memory events that are due to replication. This assumption makes the calculations analytically tractable in some cases, yet still retains the ability to explain experimental data. For a simple model of fission yeast heterochromatin, we obtained analytic expressions for replication-driven switching rates (Section \ref{fission}). For a more complex model of budding yeast heterochromatin, we successfully compared numerical results from our framework with experimental switching rate measurements (Section \ref{quantcomp}).

In the most general terms, this framework requires knowledge of the fixed points of a dynamical model for histones PTMs, and the basins of attraction of the stable fixed points. It also requires knowledge of the histone PTM state of the newly synthesised nucleosomes incorporated into chromatin after DNA replication. These considerations fully specify the probability distribution for histone PTM levels after replication and the domain of integration (the basin of attraction of the opposite fixed point), enabling the computation of the replication-driven switching rate. In general, these integrals can be difficult to compute. However, if the region of the integration domain where the Normal distribution is significant is small enough, the separatrix can be approximated by a straight line, implying that, for many practical purposes, \cref{switch_general} could have a wide applicability.

From a conceptual perspective, this work helps to establish a classification of epigenetic switching events, dissecting them by the noise source that dominates the switching behaviour: whether it is the intrinsic noise of biochemical reactions or the dilution of epigenetic marks at replication. The fact that we are only considering switching induced by replication, which is only a subset of all loss-of-memory events that could occur in cellular lineages,  is a limitation of this work. In Refs. \cite{jost2014bifurcation,micheelsen2010theory} the effects of biochemical noise have been examined, but the combination of the two sources of noise in epigenetic systems remains unexplored analytically. Given that their combined effects can be synergistic (rather than additive), further analysis of both sources of noise together will be an important avenue for future research.


\subsubsection*{Acknowledgements}
We thank Govind Menon for fruitful discussions. We also thank the BBSRC Institute Strategic Programme (BB/P013511/1) and the Wellcome Human Developmental Biology Initiative for funding.

\begin{appendices}
\renewcommand\thefigure{\thesection.\arabic{figure}}    
\setcounter{figure}{0}    

\section{Fixed points and stability analysis} \label{LSA}

Making use of the symmetry of the baseline, we can easily compute the fixed points, denoted by $\ast$. Adding and subtracting \cref{eqa0,eqm0} at steady state, with $c_+=c_a^\ast+c_m^\ast$ and $c_-=c_a^\ast-c_m^\ast$, we have
\begin{align}
(1-c_+)(2+k_c c_+)-k_d c_+ - 2 k_d k_c c_a^\ast c_m^\ast=&0 \label{c+eq} \\
(1-c_+)k_c c_--k_d c_-=&0 \label{c-eq}.  
\end{align}
From \cref{c-eq}, we have that either $c_-=0$ or $c_+=1-k_d/k_c$. The case $c_-=0$ yields the first fixed point:
\begin{equation}
c^{*}_{a,1}=c^{*}_{m,1}=\frac{1}{k_d+2}.
\end{equation}
The other case, $c_+=1-k_d/k_c$, together with the fact  $c_a^\ast c_m^\ast =(c_+^2-c_-^2)/4$ and \cref{c+eq}, yields
\begin{equation}
0=\frac{k_d}{k_c}(2+k_c-k_d)-k_d\left(1-\frac{k_d}{k_c}\right)-k_dk_c\frac{\left(1-\frac{k_d}{k_c}\right)^2-c_-^2}{2},
\end{equation}
whose solution for $c_-$ is
\begin{equation}
\label{c-app}
c_-=\pm \frac{1}{k_c} \sqrt{\left(k_c-k_d\right)^2-4}.
\end{equation}
With a slight abuse of notation, we define $c_-$ as the positive solution to the previous equation and the sign is made explicit. Then, the two solutions of \cref{c-app}, together with $c_+=1-k_d/k_c$ specify the second and third fixed points of the system:
\begin{align}
c^{*}_{a,2}=&(c_+ + c_-)\frac{1}{2}, \quad c^{*}_{m,2}=(c_+ - c_-)\frac{1}{2}  \\
c^{*}_{a,3}=&(c_+ - c_-)\frac{1}{2}, \quad c^{*}_{m,3}=(c_+ + c_-)\frac{1}{2} .
\end{align}

The local stability of the fixed points can be found with a linear stability analysis. This involves obtaining the Jacobian matrix of the dynamical system, $J_{ij}=\frac{\partial f_i}{\partial c_j}\vert_*$, where $f_i$ are the right-hand sides of \cref{eqa0,eqm0} and $\vert_*$ means that they are evaluated at the fixed point of interest. The Jacobian matrix takes the form
\begin{equation}
J=
\begin{pmatrix}
    -1-k_d -k_c[-1+2c_a +c_m(1+k_d)] &  -1-k_c c_a (1+k_d) \\
    -1-k_c c_m(1+k_d) & -1-k_d -k_c[-1+2c_m +c_a(1+k_d)]
\end{pmatrix}\Bigg \vert_\ast
\end{equation}
whose eigenvalues at the central fixed point [defined by \cref{ss01}]  are
\begin{align}
\lambda_1=-\frac{k_d (2-k_c+k_d)}{2+k_d} \nonumber \\ 
\lambda_2=-2-k_c-k_d. \nonumber
\end{align}
The associated eigenvectors are
\begin{align}
v_1=(-1,1) \nonumber \\ 
v_2=(1,1), \nonumber
\end{align}
implying that along the $(1,1)$ direction the fixed point is always stable, but along the $(-1,1)$ direction it is only stable if $k_c<2+k_d$ (note that all constants are greater than 0). Thus, \cref{ss01} is a stable fixed point if $k_c<2+k_d$ and a saddle point otherwise.

\section{Switching rate integrals} \label{Integrals}

The most general case of the integrals considered in Section \ref{switching} is the integral of the Normal distribution on one side of the line $y=\alpha x + \beta$. In this scenario, the switching rate integral becomes
\begin{equation}
S\simeq \int_{-\infty}^{+\infty} dx \int_{-\infty}^{\alpha x +\beta} dy \, \frac{2N}{\pi \sqrt{c_{a,2}^\ast c_{m,2}^\ast }}\,\textrm{exp}\left[ -2N \frac{(y-c_{a,2}^\ast/2)^2}{c_{a,2}^\ast} -2N \frac{(x-c_{m,2}^\ast/2)^2}{c_{m,2}^\ast} \right].
\end{equation}

For the first integral, we need the change of variable $u=\sqrt{\frac{4N}{c_{a,2}^\ast}} (y-c_{a,2}^\ast/2)$, and $du=\sqrt{\frac{4N}{c_{a,2}^\ast}} dy$. Then,
\begin{align}
S \simeq &\int_{-\infty}^{+\infty} dx \int_{-\infty}^{(\alpha x +\beta - c_{a,2}^\ast/2)\sqrt{\frac{4N}{c_{a,2}^\ast}} } du \,\frac{\sqrt{N}}{\pi \sqrt{c_{m,2}^\ast}}\, \textrm{exp} \left[  -\frac{u^2}{2}  -2N \frac{(x-c_{m,2}^\ast/2)^2}{c_{m,2}^*} \right] \nonumber \\ =&\int_{-\infty}^{+\infty} dx \, \sqrt{\frac{N}{2 \pi c_{m,2}^\ast}} \, \textrm{exp} \left[ -2N \frac{(x-c_{m,2}^\ast/2)^2}{c_{m,2}^\ast} \right] \left(1+ \textrm{erf} \left[ (\alpha x +\beta - c_{a,2}^\ast/2)\sqrt{\frac{2N}{c_{a,2}^\ast}} \right]\right).
\end{align}

From Ref. \cite{ng1969table}, we have that 
\begin{equation}
\int_{-\infty}^{+\infty} d\tilde{x} \, \textrm{erf} (\tilde{x}) e^{-(a\tilde{x}+b)^2}= -  \frac{\sqrt{\pi}}{a} \textrm{erf}\left( \frac{b}{\sqrt{a^2+1}} \right).
\end{equation}
In this case, $a=\frac{1}{\alpha}\sqrt{\frac{c_{a,2}^\ast}{c_{m,2}^\ast}}$ and $b=(c_{a,2}^\ast/2-\beta)\sqrt{\frac{2N}{c_{m,2}^\ast}}\frac{1}{\alpha}-\sqrt{\frac{N c_{m,2}^\ast}{2}}$. Together with $\tilde{x}=(\alpha x +\beta - c_{a,2}^\ast/2)\sqrt{\frac{2N}{c_{a,2}^\ast}}$ we have 
\begin{align}
S \simeq & \,\frac{1}{2}+\int_{-\infty}^{+\infty}d \tilde{x} \, \sqrt{\frac{c_{a,2}^\ast}{\pi c_{m,2}^\ast}}\frac{1}{2\alpha} e^{-(a\tilde{x}+b)^2} \textrm{erf} (\tilde{x}) \nonumber\\ =& \, \frac{1}{2} - \frac{1}{2} \textrm{erf} \left( \frac{(c_{a,2}^\ast /2-\beta)\sqrt{\frac{2N}{c_{m,2}^\ast}}\frac{1}{\alpha}-\sqrt{\frac{N c_{m,2}^\ast}{2}}}{\sqrt{\frac{c_{a,2}^\ast}{c_{m,2}^\ast \alpha^2}+1}}\right) \nonumber \\ =& \frac{1}{2} \left[ 1- \textrm{erf} \left( \sqrt{\frac{N}{2}}\frac{c_{a,2}^\ast-2 \beta -\alpha c_{m,2}^\ast}{\sqrt{c_{a,2}^\ast + \alpha^2 c_{m,2}^\ast}}\right) \right],
\end{align}  
which corresponds to \cref{switch_general} in the main text.

For the symmetric baseline, where the separatrix is just $y=x$ (i.e. $\alpha=1$ and $\beta=0$), we have the result stated in the main text, \cref{switching_rate} :
\begin{equation}
S \simeq \frac{1}{2} \left[ 1- \textrm{erf} \left( \sqrt{\frac{N}{2}}\frac{c_{a,2}^\ast - c_{m,2}^\ast}{\sqrt{c_{a,2}^\ast + c_{m,2}^\ast}}\right) \right]= \frac{1}{2} \left[ 1- \textrm{erf} \left( \sqrt{\frac{N}{2}}\frac{c_-}{\sqrt{c_+}}\right) \right].
\end{equation}

\section{Derivation of the budding yeast model} \label{App_SIR}

Our starting point is the stochastic model for Sir-protein silencing proposed in Ref. \cite{Miangolarra2024twoway}. With respect to that model, we neglect the methylation mark, which is not part of any feedback loop, and  we take the deterministic limit, i.e., $u=\langle n_u/N \rangle$, where $n_u$ is the discrete number of unmodified nucleosomes, and neglect correlations. We arrive at the corresponding dynamical equations:
\begin{align}
\frac{\textrm{d} s}{\textrm{d} t}=& -s+\frac{u^2 c_\textrm{SIR} k_D}{2V} \label{eqs}\\
\frac{\textrm{d} u}{\textrm{d} t}=&  s-\frac{u^2 c_\textrm{SIR} k_D}{2V}-k_{sas2} u  +k_{sil} a + a s \frac{k_{sir2}}{V} \\
\frac{\textrm{d} a}{\textrm{d} t}=& k_{sas2} u  -k_{sil} a - a s \frac{k_{sir2}}{V}, 
\end{align}
where $a$ is the fraction of acetylated nucleosomes,  $u$ the fraction of unmodified nucleosomes, $s$ the fraction of Sir-bound nucleosomes and $a+u+s=1$. In these equations the unbinding rate of the Sir complex has been absorbed by a rescaling of time ($t$). The Sir proteins dimerise to bind to nucleosomes \cite{Behrouzi2016heterochromatin}, hence giving the term $u^2 c_\textrm{SIR} k_D/(2V)$, where $k_D$ is the characteristic constant of the binding by dimerisation reaction and $V$ is the effective volume occupied by the locus. The volume dependency arises since a smaller volume would  introduce more contacts and would therefore increase the dimerisation rate. $k_{sas2}$ is the acetylation rate constant, $k_{sil}$ the silencer-mediated deacetylation rate constant (this term has also been simplified with respect to the full model of Ref. \cite{Miangolarra2024twoway}), and $k_{sir2}$ the Sir-mediated deacetylation rate constant. This last term depends on the fraction of Sir-bound nucleosomes and their contacts with acetylated nucleosomes, which is why it is also inversely proportional to the volume of the locus. The volume of the locus, for practical purposes, can be approximated by
\begin{equation}
V=\frac{V_{max}}{1+(V_{max}-1)(u+s)},
\end{equation}
where $V_{max}$ is the ratio between the largest and the smallest volume occupied by  the locus and is therefore a dimensionless parameter. The changes in the locus volume (or, conversely, changes in chromatin compaction) depend on the number of deacetylated nucleosomes, as found in many studies \cite{Collepardo2015chromatin,Robinson2008_30nmAc,Valenzuela2008long} and as argued  in Ref. \cite{Miangolarra2024twoway}.

Given that the binding and unbinding of Sir proteins occurs at a much shorter timescale \cite{Behrouzi2016heterochromatin} than the acetylation dynamics \cite{Waterborg2001}, we can assume that \cref{eqs} is always at steady state, obtaining
\begin{equation}
\label{eqsss}
s^\ast(u)=\frac{u^2 k_D c_\textrm{SIR} (u-1-V_{max} u)}{u^2 k_D c_\textrm{SIR} (V_{max}-1)-2V_{max}}.
\end{equation}
This, together with the constraint $a+u+s=1$,  reduces the dynamical system to a single dimension:
\begin{equation}
\label{simp_model}
\frac{\textrm{d} u}{\textrm{d} t}=-k_{sas2} u  +k_{sil} [1-u-s^\ast(u)] + \frac{[1-u-s^\ast(u)] s^\ast(u) k_{sir2}}{\frac{V_{max}}{1+(V_{max}-1)[u+s^\ast(u)]}},
\end{equation}
where $s^\ast(u)$ is given by \cref{eqsss}. This is the form of the model given in the main text.

\begin{figure}
\center
\includegraphics[width=0.65\textwidth]{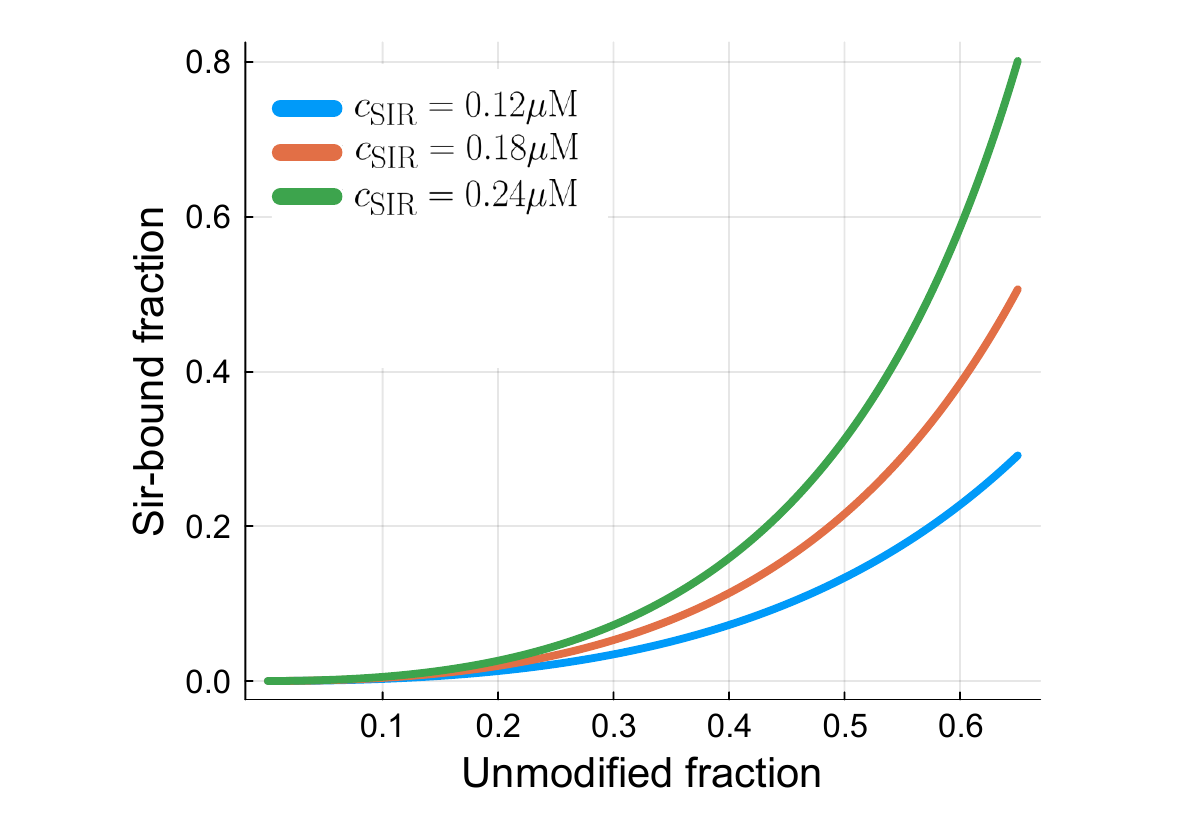}
\caption{$s^\ast(u)$  as a function of $u$, as given by \cref{eqsss_mt}, for different values of $c_\textrm{SIR}$. The rest of the parameters are as specified in Appendix \ref{budding_numerics}. It can be seen that the relationship is monotonic in the range depicted here, which is the relevant range, since we require $0\le u \le 1$, $0\le s^\ast(u) \le 1$ and $u+s^\ast(u) \le 1$.  \label{FigA1}}
\end{figure}

As noted in the main text, $s^\ast(u)$  is monotonically increasing with $u$. In Fig. \ref{FigA1}, we show this relationship for the parameters used in Fig. \ref{Fig4} and for various values of $c_\textrm{SIR}$.

\section{Numerical solution and fit of the budding yeast model} \label{budding_numerics}

To obtain Fig. \ref{Fig4}A, for each set of parameters, the fixed points of \cref{eq_csir} need to be obtained (shown in Fig. \ref{Fig4}B), which are then fed into \cref{SIR_switching} to obtain the replication driven switching rates. The curve in Fig.  \ref{Fig4}A is obtained by varying $c_\textrm{SIR}$ from 0.05 to 0.4 $\mu M$.

For the fitting procedure that determined the parameters used for Fig. \ref{Fig4}A, we took a manual approach because the parameters cannot be varied freely, as that could result in a loss of bistability and, if the system is not bistable, the switching rates are ill-defined. We defined an $R^2$ cost-function to determine the goodness of fit between the model and the experimental data. Manually varying the parameters (within the bistable region) we approximately located a local minimum of the cost function. The parameter values resulting from this procedure were $k_{sir2}=24.5\pm 0.1$ and $k_{sas2}=0.8\pm 0.01$. The values of the rest of parameters where fixed to $V_{max}=1.5^3$ (as in Ref. \cite{Miangolarra2024twoway}), $k_{sil}=0.02$ and $k_D=12 \mu M^{-1}$ (as in Ref. \cite{Miangolarra2024twoway}). $k_{sir2}$, $k_{sas2}$ and $k_{sil}$ are dimensionless (originally their units would be the inverse of time, but time has been renormalised and non-dimensionalised). The ratio between the values obtained for $k_{sir2}$ and $k_{sas2}$ are of the same order of magnitude as the ratio of their original values in Ref. \cite{Miangolarra2024twoway} (30.6 vs 16.8 in the original paper).

For more details, the code is publicly available at \url{https://github.com/AMovillaMiangolarra/replication_driven_switch_aux_code}.

\end{appendices}

\bibliographystyle{siam}
\bibliography{biblio}

\end{document}